# Composite nanowires for room-temperature mechanical and electrical bonding

Yanbin Cui [a, b] and Yang Ju [b]

[a] State key Laboratory of Multiphase Complex System, Institute of Process Engineering, Chinese Academy of Sciences, Beijing 100190, China

[b] Department of Mechanical Science and Engineering, Nagoya University, Nagoya 464-8603, Japan

**Contents**

**Abstract:**

At millimeter dimension or less, the conventional bonding technology in electronic assembly relies heavily on reflow soldering and suffers from severe performance and reliability degradation. Meanwhile, the traditional high temperature bonding process (easily reach 220 $^{o}$C) tends to result in undesired thermal damage and residual stress at the bonding interface. It is therefore a major challenge to find a means to preparing room-temperature connectors or fasteners with good mechanical and electrical bonding. Very recently, composite nanowire have been used to fabricate room-temperature fasteners. In this chapter, we summarize the state-of-the-art progress on the use of composite nanowires for room-temperature mechanical and electrical bonding. Using anodic aluminum oxide (AAO) and polycarbonate (PC) membrane as template, the fabrication of Cu/parylene and Cu/polystyrene nanowires were described. Meanwhile, the fabrication of carbon nanotube (CNTs) array was summarized. Then, the performances of the composite nanowires (Cu/parylene, Cu/polystyrene and CNT-Cu/parylene) used as surface fastener for room-temperature mechanical and electrical bonding were demonstrated.

# 1. Introduction

Surface mount devices (SMDs) relies heavily on reflow soldering and has become the cornerstone of today's electronic industry. The heating temperatures of traditional reflow soldering technique can easily reach 220 °C during reflow soldering, which may cause not only energy consumption but also thermal damage to the surface mount components. Additionally, the toxicity of traditional Sn-Pb solder has led to a trend of worldwide legislation that mandates the removal of lead from electronics. Although various types of lead-free solder have been proposed and adopted in the electronics industry, the melting points of these lead-free solders are always 5-20 °C higher than Sn-Pb solder. Moreover, the recycling of rare metals in the surface mount components and printed circuit boards is not easy due to the difficulties in detaching the components, circuit boards and solder materials. [1] On the other hand, the continuous trend toward miniaturization and functional density enhancement makes urgent the demand to improve the bonding technology in surface mount technology (SMT). At millimeter dimensions or less, conventional electrical connectors or fasteners tend to suffer from severe performance and reliability degradation. [2] It is therefore a major challenge to find a nontoxic and room-temperature bonding technique to afford good mechanical bonding as well as electrical contact, especially for micro/nano-electronic circuits and flexible electronic devices.

With regard to developing a room-temperature bonding technique for SMT, one possible approach is to make use of cold welding. Cold welding of thin gold films on elastomeric supports has been carried out under ambient conditions and low loads. However, only the lower limit of approximately 0.1 N/cm$^2$ was reported for the adhesion strength. [3] Besides, many researchers have also succeeded in joining individual nanostructures by nanoscale-welding method. [4-6] Although direct heating was not performed and large forces were not applied in these nanoscale-welding techniques, fine manipulation of an individual nanowire or nanotube by specific equipment was always needed. Therefore, these nanoscale-welding techniques, in which the connection of two nanowires or nanotubes was performed, are suitable for nanoscale connection but inefficient for the mass production of SMDs.

Recently, hybrid core-shell nanowire forests have been used to fabricate electrical and chemical connectors. [7, 8] Specifically, we prepared a serial of nanowire (including metallic and hybrid core-shell nanowire arrays) and carbon nanotube (CNT) arrays, in which the nanowire and CNT arrays were used as a fastener for SMT. [1, 2, 9-13] For SMT, the fastener should ideally have both high adhesion strength and provide a good electrical connection. In this chapter, we summarize the state-of-the-art progress on the use of composite nanowires for room-temperature mechanical and electrical bonding. First, the fabrication of anodic aluminum oxide (AAO) membrane, which used for the growing Cu nanowires, was presented. Then, using AAO and polycarbonate (PC) membrane as template, the synthesis of copper/parylene and copper/polystyrene composite nanowires were described. Meanwhile, the fabrication of CNT arrays was also summarized. Lastly, the performances of the composite nanowires used as surface fastener for room-temperature mechanical and electrical bonding were demonstrated.

## 2. Fabrication of anodic aluminum oxide membrane

Anodic aluminum oxide (AAO) nanoporous membrane is popular for its self-organized nanostructure. Due to its highly ordered porous structure, significant thermal stability and cost effectiveness. [14, 15], AAO membranes are widely used for the fabrication of nanowires and nanotube arrays, in filtration, as sensors, catalysts, in solar cells etc. [14, 16-19] Since Masuda *et al.* [20] introduced two-step anodization in 1995, the AAO got more attraction in the field of nanotechnology. In order to produce highly ordered nanopore structures, a lot of researches had been done in terms of process parameters. By changing anodizing parameters (such as electrolytes, anodizing voltage, anodizing time, temperature and etching methods), the structure of AAO (such as pore size, pore depth, inter-pore distance, thickness of membrane and pore geometry) can be easily controlled. [21-24] AAO templates exhibit columnar pore structure, vertical to the substrate and parallel to each other with pore diameters from several tens to hundreds nanometers and with an aspect ratio between 10-1000 or more. [21, 25, 26] By filling the pores of the AAO templates, arrays of well-aligned nanowires or other 1D nanostructures with uniform diameter and length can be obtained using electroplating or other growth methods. [27]

In general, two steps anodization method was employed to produce well-arranged porous AAO template. [20] Typical fabrication processes of the AAO template are shown in Figure 1. Appropriate electrolyte solution (such as oxalic acid) was used for the anodization. The pretreated aluminum foils were anodized in electrolyte solution under constant voltage at room temperature. At the first anodization step, the alumina layer fabricated was dissolved by wet chemical etching in a mixture of 6 wt.% phosphoric acid and 1.5 wt.% chromic acid solution for 30 min at 60 °C. The resulting inner aluminum foils have uniform concave nanoarray, which is crucial to achieving ordered pore size distribution. [28] After removal, the second anodization step was carried out at the same condition. After two steps anodization process, the remaining aluminum on the AAO was dissolved in the solution made up of one part by volume of 0.1 mol/L $CuCl_2$ solution and four parts by volume of 10 wt.% HCl. [28] At last, the oxide barrier layer of the AAO was removed and the pores were enlarged by floated the sample on the surface of 0.5 mol/L phosphoric acid solution at room temperature owing to the surface tension of the AAO. [29]

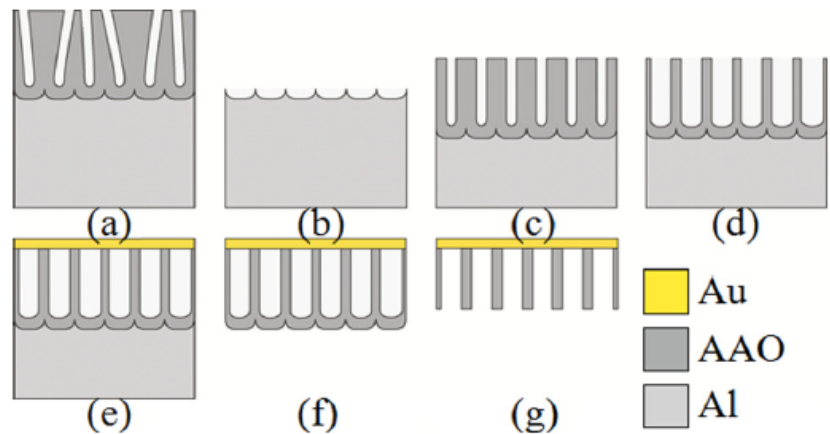

**Figure 1.** Schematic of fabricating the AAO: (a) a first step anodization, (b) wet etching of AAO fabricated in a first step anodization, (c) a second step anodization, (d) extending the pores, (e) sputtering Au film, (f) dissolving aluminium substrate, and (g) dissolving barrier layer and extending pores. [10]

It is well known that the value of the anodizing voltage for preparing the best nanopore order in self-ordered AAO is different for each electrolyte. [30] For instant, in sulfulic acid, oxalic acid and phosphoric acid, anodizing voltage is 25, 40 and 195 V, respectively. If a higher anodizing voltage is applied, the higher current density and Joule heat generation will cause damage or collapse of the pore structure. In practice, a higher potential is beneficial in increasing the pore size. Chung *et al.* proposed a high-potential hybrid pulse anodization (HPA) technique to resolve this problem. In HPA technique, a period of small negative potential is applied to suppress the Joule heating effect during the AAO preparation process. The scanning electron microscope (SEM) results showed that HPA with an anodizing potential of 60 V resulted in an intact pore structure on the AAO surface (Figure 2). By contrast, the AAO formed using conventional direct current anodization (DCA) with the same anodizing potential contained many small irregular pores around each original pore [Figure 2(a)]. [31] On the other hand, it is clearly seen that the reduction of the irregular small branch pores in AAO formed by HPA, as shown in Figure 2(b). Chung *et al.* also found that the HPA technique not only merits manufacturing convenience and cost reduction but also promotes pore distribution uniformity of AAO at severe conditions of low-purity Al foils and relatively high room temperature. The pore distribution uniformity can be improved by HPA compared with the DCA. Very good AAO distribution uniformity (91%) was achieved in high-purity aluminum foil by HPA because it can suppress the Joule's heat to diminish the dissolution reaction. [32]

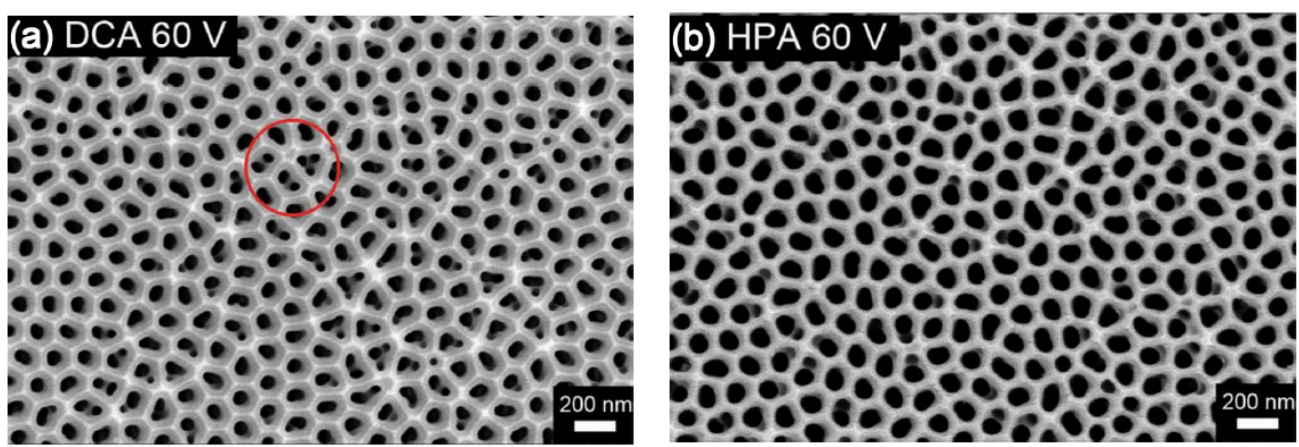

**Figure 2.** Top-view SEM micrographs of AAO nanostructures formed by (a) DCA 60 V, (b) HPA 60 V with following pore widening for 10 min. [31]

Controlling the size and uniformity of the pores is key in manipulating the structure and properties of the materials confined within the pores of AAO. Kim *et al.* reported a new approach for preparing pretextured surface patterns on aluminum using solvent-dependent differential swelling of block copolymers. Long-range order in a hexagonal-packed monolayer was created using solvent-induced ordering of poly(styrene-b-4-vinylpyridine) (PS-b-P4VP) micellar films, followed by surface reconstruction to make nanoporous patterns. Using reactive ion etching (RIE), patterns were transferred to the Al surface. Subsequent anodization in concentrated sulfuric acid enabled the formation of channels with long-range lateral order. Highly ordered porous alumina with a hole interval of 45 nm and a hole size of 12 nm was produced (Figure 3). [33]

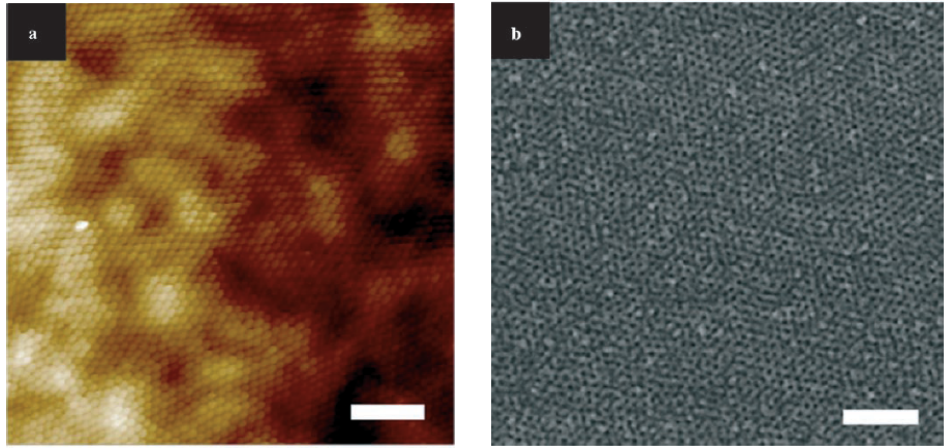

**Figure 3**. Scanning force microscopy (SFM) image of surface view of anodized Al in 2.5M sulfuric acid at 19 V at 48 °C for 10 min (a) and SEM image of anodized Al after pore widening carried out in 5 wt% phosphoric acid at 20 °C for 5 min (b). (Scale bar: 200 nm.) [33]

Besides, ultrathin AAO membranes were also be used as masks for the fabrication of nanoparticle arrays on different surfaces. The AAO membranes can be formed directly on some surfaces, such as silicon and indium tin oxide. Typically, a thin layer of Al is deposited on these substrates, and nanoporous masks are fabricated by the anodization of this layer. [34, 35] Pastore *et al.* fabricated ultrathin AAO membranes by applying low anodization voltages providing low reaction speed during AAO formation. A low anodization speed is required to get membrane thickness below 100 nm in a reproducible way. With this procedure, AAO membranes with pore diameters below 20 nm and membrane thickness below 70 nm were obtained. [36]

## 3. Synthesis of copper/parylene composite nanowires

Due to its ability to control nanomaterial shape, size and uniformity, template-assisted synthesis has attracted considerable attention and proved to be a simple and versatile approach for preparing ordered nanowire, nanorod, nanoparticle and nanodot arrays in a wide range of materials. [37, 38] Numerous nanomaterials have been prepared by utilizing hard templates such as track-etched polycarbonate (PC) membranes, [39] AAO templates [40, 41] and soft templates such as polymers [42] and surfactants. [43] On the other hand, various techniques (such as chemical vapor deposition, [44] molecular beam epitaxy, [45] vapor-liquid-solid growth process [46] and hydrothermal process) [47] have been applied to synthesize nanowire- and/or nanotube-structured materials. Compared to those methods, electro-deposition is one the most cost-effective techniques to fabricate the nanostructured materials. [48] Therefore, directional growth of nanowires/rods/pillars through PC or AAO templates via electro-deposition process is very popular method because of its inexpensive, simple technique and ability to incorporate complex geometrical structures, easy tenability of the nanopores' dimensions (so also nanowires/rods' dimension) by controlling the deposition parameters. [49]

Recently, Ge/parylene core/shell nanowire array was used as an electrical connector after the deposition of an Ag film with relatively high shear adhesion strength. [8] In addition, Ni [50] and Cu [51] nanowire arrays have been found to have very low electrical resistance. We also prepared nanowire surface fasteners based on gold and copper nanowire arrays, [1, 9] with relatively low electrical resistance and adhesion strengths. The maximum adhesion strength of the metallic nanowire surface fastener is only 8.17 N/cm$^2$. Therefore, it is still a challenge to achieve high adhesive strength and low electrical resistance at the same time for the room-temperature surface fastener. In order to improve the performance, we fabricated copper/parylene core/shell nanowire surface fastener. [11, 12] Compared with metallic nanowire surface fastener, the adhesion strength increased dramatically.

A peculiar cell (Figure 4) was used to fabricate freestanding copper nanowire on the substrate directly. The introduced porous glass and porous cellulose membrane have three important functions. First, the capillary forces provided by the porous glass plate and porous cellulose membrane help maintain a continuous electrolyte flow from the bulk of the electrolyte to PC membrane. [52] Second, the stiffness of porous glass plate ensures the contact of the substrate with PC membrane. Third, the compliance of porous cellulose membrane offers a buffer and then ensures tight contact of the substrate with PC membrane. It is known that for flat anodes the metal is deposited preferentially at the outer border areas of the cathode. [12] This effect was avoided by using a conical copper anode, leading to a noticeably more homogeneous copper nanowire distribution over the whole cathode surface. [53] Before and after the assembly of the cell, two additional immersions were introduced to ensure an even copper ion density throughout the PC membrane (Millipore ISOPORE). [9] Copper nanowire arrays were then synthesized by electro-deposition under a constant current. The electro-deposition electrolyte used was a 0.4 M $CuSO_4·5H_2O$ solution, adjusted to pH 2 with sulfuric acid. The electro-deposition was performed at room temperature and without stir. After etching in methylene chloride to remove the PC membrane, the freestanding copper nanowire arrays on the substrate were obtained. Then, a thin film of Parylene C was deposited on copper nanowire arrays by using a DACS-LAB deposition system. The typical deposition conditions were 160 °C for the evaporation of the parylene dimer precursor, 650 °C for the pyrogenic decomposition of the dimer into monomers, and 60 mTorr for the vacuum chamber. Through controlling the amount of the loaded precursor, the corresponding thickness of parylene shell was obtained.

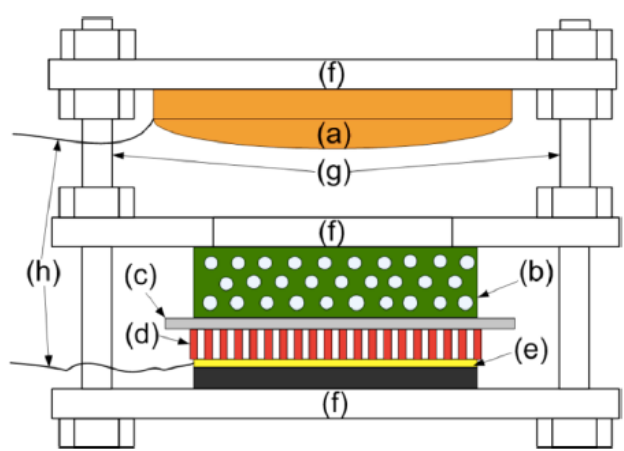

**Figure 4.** Schematic of the cell for copper nanowire fabrication: (a) conical copper anode, (b) porous glass plate, (c) cellulose membrane, (d) polycarbonate template, (e) glass substrate with gold film, (f) isolation holder, (g) screw and nut, and (h) copper wire. [12]

The fabrication procedure of the copper/parylene core/shell nanowire surface fastener is outlined in Figure 5(a). At first, copper nanowires were grown on glass/Cr/Au substrates by the template-assisted electro-deposition method. After etching the PC template, a thin layer of Parylene C was evenly deposited on the copper nanowires to enhance the adhesive ability of nanowire surface fastener. The SEM image of the copper nanowire arrays with an average diameter of 150 nm [Figure 5(b)] indicates that most of the nanowires were grown vertically on the substrate but oriented in a wide range of directions. Figure 5(c) and (d) show the SEM image of copper nanowires with a 100 and 200 nm parylene coating, respectively. Clearly, the grown copper nanowires sustain their high aspect ratio without aggregation, partly due to the high Young's modulus of the copper (~110 GPa).

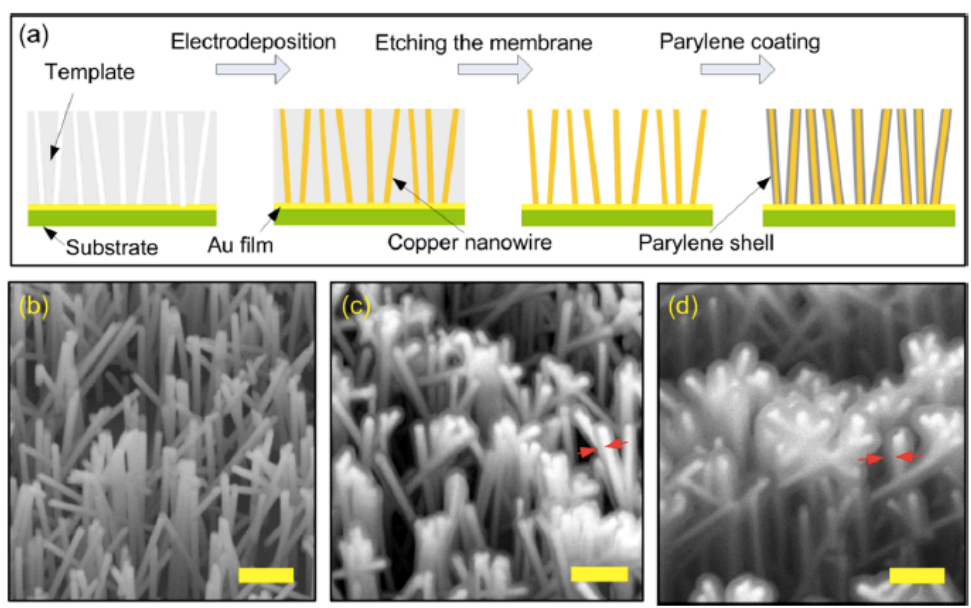

**Figure 5.** (a) The fabrication process of copper/parylene core/shell nanowire. SEM images of copper nanowires (b) without parylene coating, (c) with 100 nm parylene coating, and (d) with 200 nm parylene coating. The red arrows in parts c and d indicate the parylene shell. The scale bar is 1 μm. [12]

## 4. Synthesis of copper/polystyrene composite nanowires

To further improve the performance, a copper/polystyrene core/shell nanowire surface fastener was prepared, which shows higher adhesion strength and a much lower electrical resistance than the copper/parylene core/shell nanowire surface fastener. [11, 54]

The fabrication procedure of the copper/polystyrene core/shell nanowire surface fastener is shown in Figure 6. The AAO membranes (Synkera Company) with a nominal pore diameter of ~80 nm and interpore distance of ~240 nm were used as templates for the synthesis of polystyrene nanotubes. These open-ended nanotubes were synthesized into the AAO template by spin-casting method. [54, 55] Briefly, a polymer solution (~20 μl of 2.5 wt% polystyrene solution in toluene) was directly loaded over the spinning AAO template. After 90 s of spinning, the toluene was evaporated and then the polystyrene nanotubes were formed in the template. The thickness of polystyrene nanotube can be controlled through adjusting the loaded times. After the formation of nanotubes, the copper nanowire array was fabricated into polystyrene nanotubes using template-assisted electro-deposition method [54] by setting up the stacked cell (Figure 4) in a 0.4 M $CuSO_4·5H_2O$ solution, under a constant current of 3 mA at room temperature. The pH value of the copper sulfate solution was maintained at pH 2 with sulfuric acid. Finally, the copper/polystyrene core/shell nanowire array was obtained after etching the AAO template by 3 M NaOH solution.

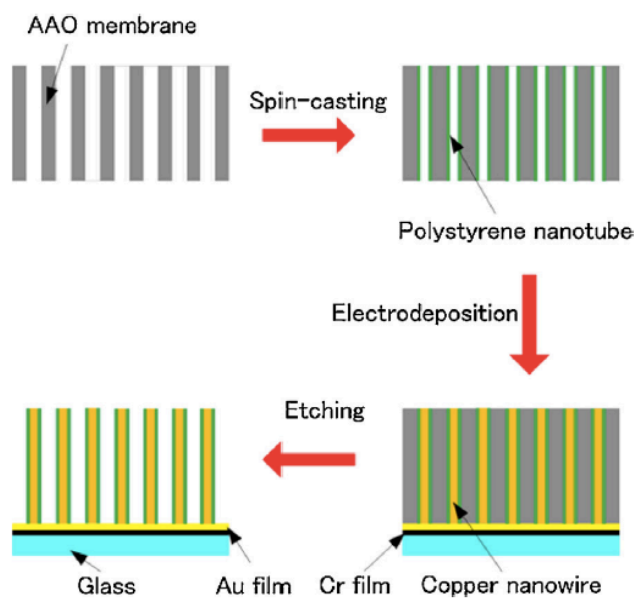

**Figure 6.** Schematics of the fabrication process for the copper/polystyrene core/shell nanowire array. [11]

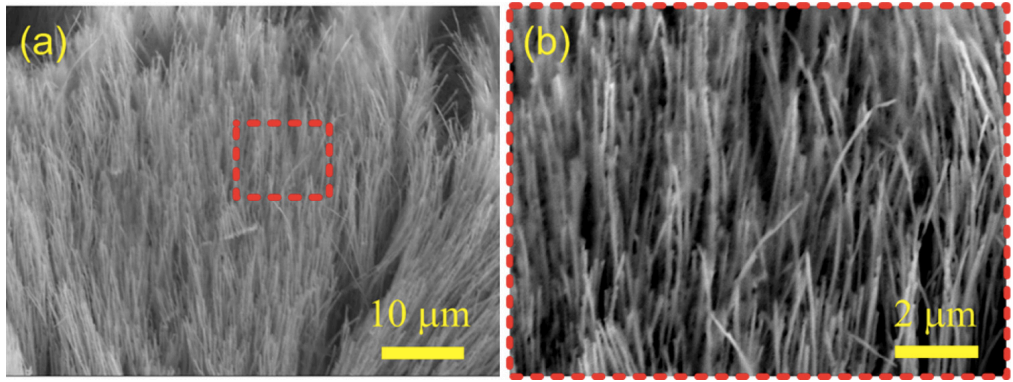

**Figure 7.** SEM image of a typical copper/polystyrene core/shell nanowire array. [11]

A SEM image of a typical copper/polystyrene core/shell nanowire array is shown in Figure 7, clearly indicating the mainly vertical orientation and the uniform shell thickness of the nanowires. Each of the components plays an essential role in achieving the desired functionality of electrical nanowire surface fasteners. The Cu core provides the following functions: (I) an electrical conductive function as the polymer shell shrinks to nanoscale; [12] (II) preventing aggregation and/or collapse of nanowires thereby resulting high aspect ratio structures. The polystyrene shell enhances the surface compliance and the adhesion energy, therefore, increasing the adhesion strength.

## 5. Fabrication of carbon nanotube array

CNTs have been among the most scientifically studied materials for the past two decades. [56] Due to their unique properties, such as extremely mechanical properties, high electrical conductivity and thermal conductivity/stability, [57, 58] CNTs have been suggested for a variety of practical applications. [59-62] The most intriguing properties of CNTs lie in their unique one-dimensional nanoscale structures that are anisotropic: properties in the longitudinal direction are drastically different from those in the azimuthal directions. [63] It is highly desirable in the realization of most of the applications to control the orientation of CNTs either as a stand-alone individual or in a group of many. [64]

CNT arrays typically comprise billions of CNTs per square centimeter on a substrate. CNTs in a CNT array are nearly parallel to each other and have more uniform direction and length distribution than other kinds of CNTs such as tangle CNTs. CNT arrays can provide a well-defined large surface area and they can be readily incorporated into devices with improved uniformities in mass production, greatly facilitating their integration into practical devices. [57] Due to highly ordered and high surface area materials with excellent electronic and mechanical properties, [65] CNT arrays have wider applications than random bulk CNTs. These applications include field emission cathodes, [66] supercapacitors, [67] nanofiltration membranes, [68] and fuel cell and solar cells. [63]

CNTs in CNT arrays are straight and uniformly aligned in the vertical direction, which is beneficial for fastener applications. Compared with metallic nanowairs, CNTs are more bendable and have significant entanglement, both of which enhance side contact with fibrillar arrays. Moreover, CNTs possess good mechanical properties and high electrical conductivity, and are consequently predicted to have potential applications in electrical fasteners. [2] Therefore, CNT and Cu/parylene core/shell nanowire array was used to construct an electrical fastener. Compared with a metallic nanowire array fastener, [1, 9] the adhesion strength of the CNT-Cu/parylene nanowire array fastener is significantly improved, offering a six-fold increase that is critical for pushing forward practical applications of room-temperature electrical fasteners. [64]

A Si wafer (<100>type, 1–10 $\Omega$/cm) with a 600 nm $SiO_2$ layer was used as a substrate. A catalyst film of Fe (0.5-1.0 nm) was deposited on the Si wafer by electron-beam (E-beam) evaporation (Edwards EB3 Electron Beam Evaporator). The evaporation was carried out at a pressure of about $5\times10^{-7}$ Torr. The deposition rate of the Fe film was kept at 0.05 nm/s to achieve uniform and controllable film thickness. The thickness of the thin Fe film was monitored in situ by a quartz-crystal sensor fixed inside the E-beam evaporation chamber and calibrated ex situ by atomic force microscopy (AFM, Veeco Explorer).

The CNT arrays were synthesized in a quartz tube furnace. The reaction chamber was a quartz tube with 3-inch diameter. The substrates were placed in the middle of the quartz tube and the quartz tube was pumped down to 5 mTorr to remove any ambient gas. The total pressure was maintained at 1 atm for all the experiments. The furnace was heated up to the growth temperature (690-780 °C) at 20 °C/min and under Ar flow. Then, $C_2H_2$ (99.999 %) and/or $H_2$ (99.9999 %) were introduced into the reactor and the CNT arrays were synthesized at the growth temperature for a certain time. Lastly, the acetylene and/or hydrogen gas flow was turned off and the furnace was cooled down to room temperature with the Ar gas flow continuing. [64]

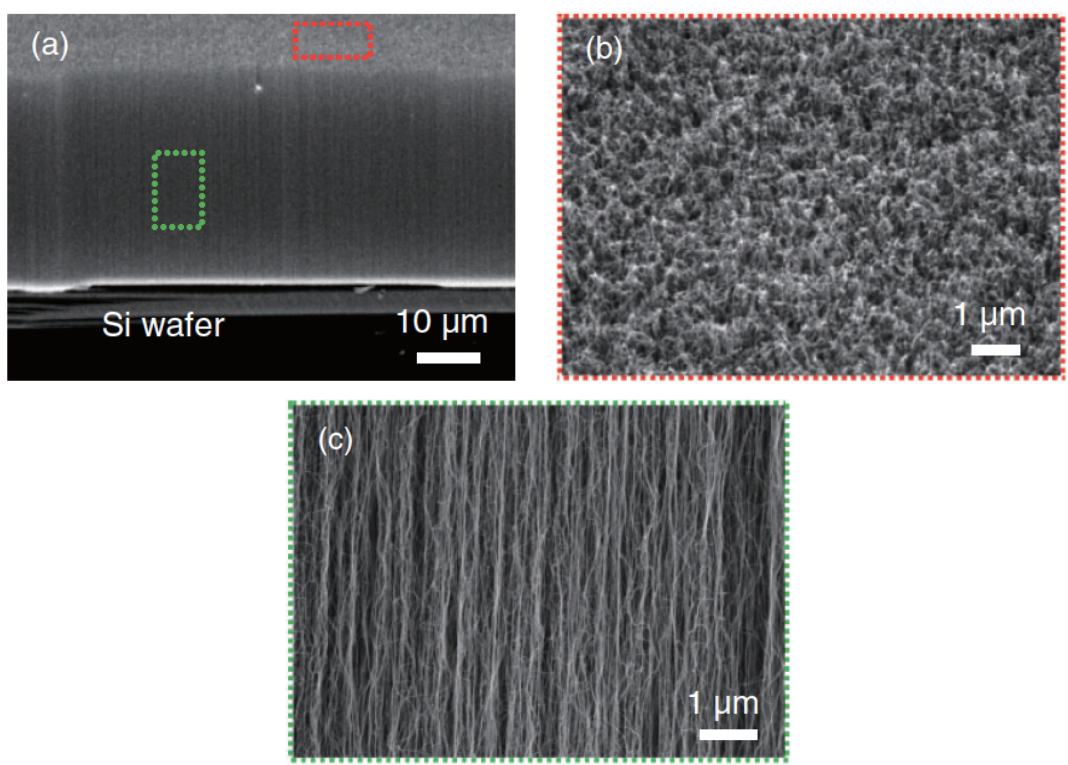

**Figure 8.** (a) Cross-sectional SEM image of CNT array. (b) Top view of CNT array showing entanglement of CNTs at the surface. (c) High magnification SEM image showing the alignment of CNTs in the array side-wall. [2]

Figure 8 shows typical SEM images of CNT arrays on a Si substrate, grown under conditions of $C_2H_2$ =100 sccm, $H_2$=50 sccm, and 750 °C. In the top layer of CNT array, it can be seen that the CNTs are tangled and not in alignment [Figure 8(b)]. It should be noted that the CNT array in this case was prepared by a normal chemical vapor deposition process, which can be easily scaled-up for the preparation of CNT arrays. The morphology of the top surface is similar to that of other CNT arrays prepared by low-pressure chemical vapor deposition methods. [69] As shown in Figure 8(c), the CNTs in the array are well aligned and bundled in the vertical direction, with these bundles being parallel to one another, while the waved CNTs switch between different straight CNT bundles. The curved CNTs in the top surface of the array, and the waved CNTs, can entangle and coil on the surface of Cu/parylene nanowires. This in turn increases the contact area between CNTs and Cu/parylene nanowires, which enhances the adhesion strength of the fastener. [2]

## 6. Mechanical and electrical performances of nanowire surface fasteners

As shown in Figure 9, a specific pattern for the fastener areas and printed wires (1 mm width) were designed to facilitate the testing of the mechanical bonding strength and the parasitic resistance of the electrical bonding. The diameter of each of the four fastener areas was 2 mm. [12] Au films approximately 100 nm thick were deposited onto the fastener and wire areas on Si or glass substrates with around 50 nm thick Cr adhesive layer to improve the adhesion between the substrate and the Au film. The adhesive layer and the Au film were deposited via E-beam evaporation or sputtering. [1, 11]

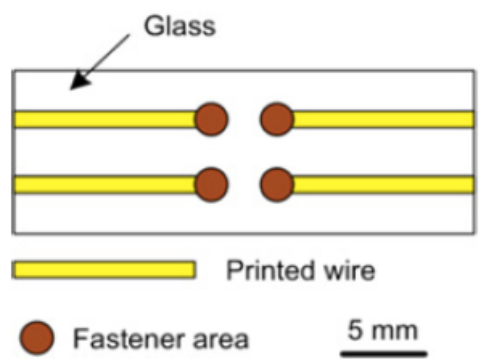

**Figure 9.** Sketch of a sample with specific pattern. [54]

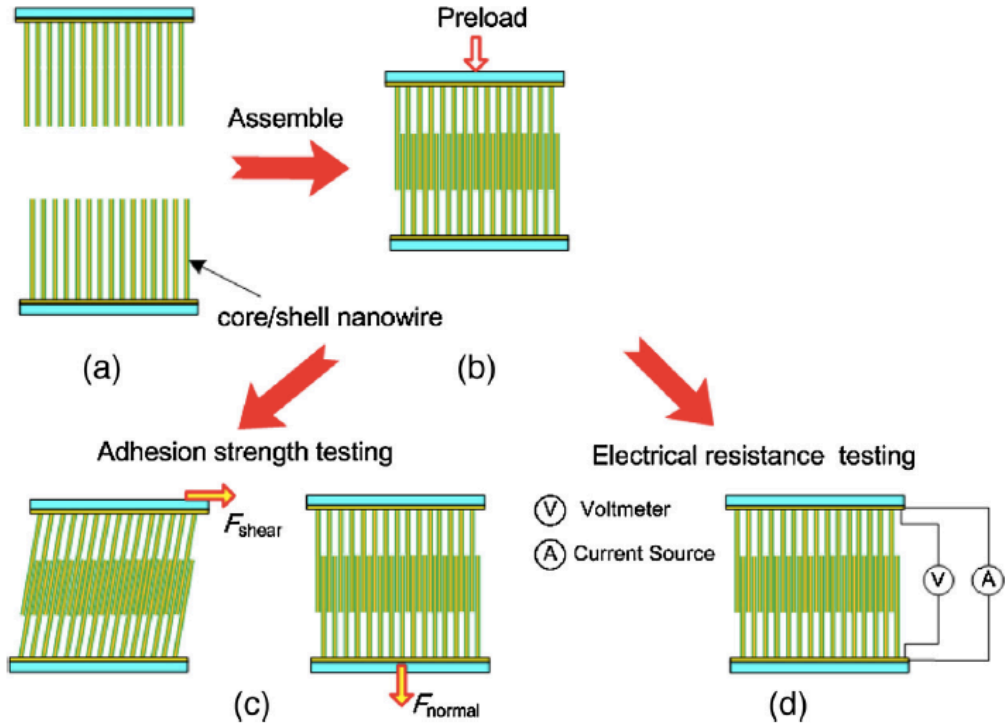

**Figure 10.** The assembly of patterned nanowire arrays as NSF, together with the adhesion strength and electrical resistance testing setup. [11]

To better understand the adhesion performance of the nanowire surface fastener, we carried out macroscopic measurement of adhesion strength, which is defined as the force acted to the nanowire surface fasteners to separate their bonding divided by the bonding area, in normal and shear directions. As shown in Figure 10, two nanowire surface fastener samples [Figure 10(a)], which had patterned nanowire arrays on the substrate, were brought into interconnection at a preload force [Figure 10(b)]. [9] After the preload force was completely released, the weight of balance was used to measure the pull-off forces with the parallel (i.e. the shear adhesive strength) and normal (i.e. the normal adhesive strength) directions to the substrate [Figure 10(c)]. [11] To eliminate the effects of the attachment cycles and avoid the pre-alignment effect for subsequent adhesion measurements, the adhesion forces obtained from the first attachment of each sample were adopted. Adhesion tests for each reported condition were completed for 4 samples, and the average values were used. The four-point probe method was used to measure the electrical resistance [Figure 10(d)]. During the measurement, an electrical current in the range from 0 to 19 mA was applied using the current source and the corresponding voltage was extracted from the voltmeter. [11, 12]

**6.1 Performances of copper/parylene nanowire surface fasteners**

To characterize the properties of copper/parylene nanowire surface fasteners, we first measured the adhesive strength and relative electrical resistance as a function of nanowire length (5, 10, and 20 μm). All the samples in this test have a parylene shell (100 nm thickness) and a preload of 78.02 N/cm$^2$. Interestingly, the Parylene C film becomes conductive due to dielectric breakdown when the thickness of it is miniaturized to nanoscale. The reason why Parylene C film in nanoscale thickness becomes conductive is mainly due to the dielectric breakdown phenomenon. [12] As can be seen in Figure 11(a), both shear and normal adhesion can be realized at the same time. Moreover, the adhesive strength is strongly affected by the length of the nanowires. The maximum shear and normal strengths were obtained when the length was 10 μm. When L < 10 μm, the nanowires sustained their high aspect ratio and the neighboring nanowires did not contact with each other [Figure 12(a)]; therefore, the contact area is directly proportional to the nanowire length. However, when the nanowire length is as large as 20 μm, the nanowires tend to collapse and the neighboring nanowires contact with each other [Figure 12(c)], which leads to a reduction of the contact area of the nanowire surface fasteners. The electrical resistance is also strongly affected by the length of nanowires. Specifically, longer nanowire length results in smaller electrical resistance [Figure 11(b)]. As can be seen in Figure 12(a)-(c), the interconnection of neighboring nanowires increases as the length of nanowires increases, and they were interconnected before the parylene coating. The interconnected neighboring nanowires connect in parallel in the electrical connection, which led to the reduction of resistance.

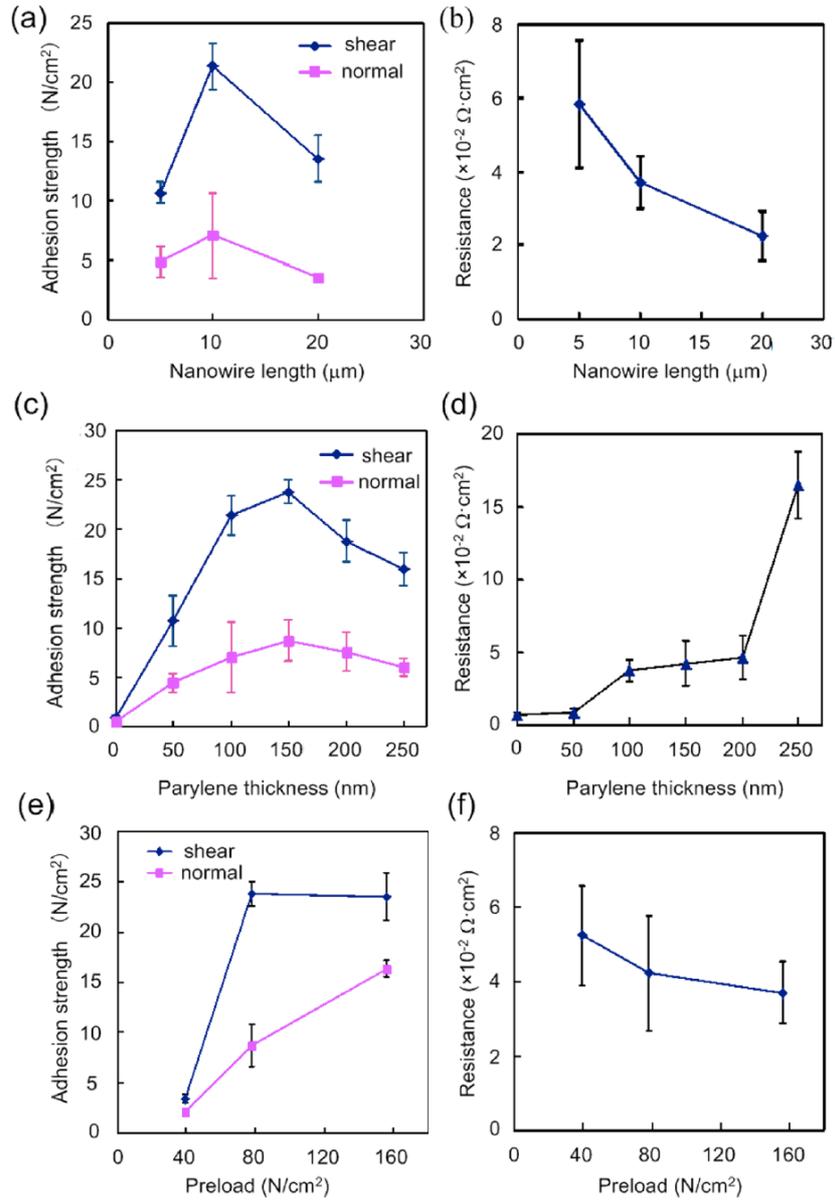

**Figure 11.** (a) Adhesive strength and (b) electrical resistance of nanowire surface fasteners as a function of nanowire length. The preload is 78.02 N/cm$^2$ and the thickness of the parylene shell is 100 nm. (c) Adhesive strength and (d) electrical resistance of nanowire surface fasteners as a function of parylene thickness. The preload is 78.02 N/cm$^2$ and the length of the nanowire array is 10 μm. (e) Adhesive strength and (f) electrical resistance of nanowire surface fasteners as a function of preload. The thickness of parylene shell is 150 nm and the length of nanowire array is 10 μm. [12]

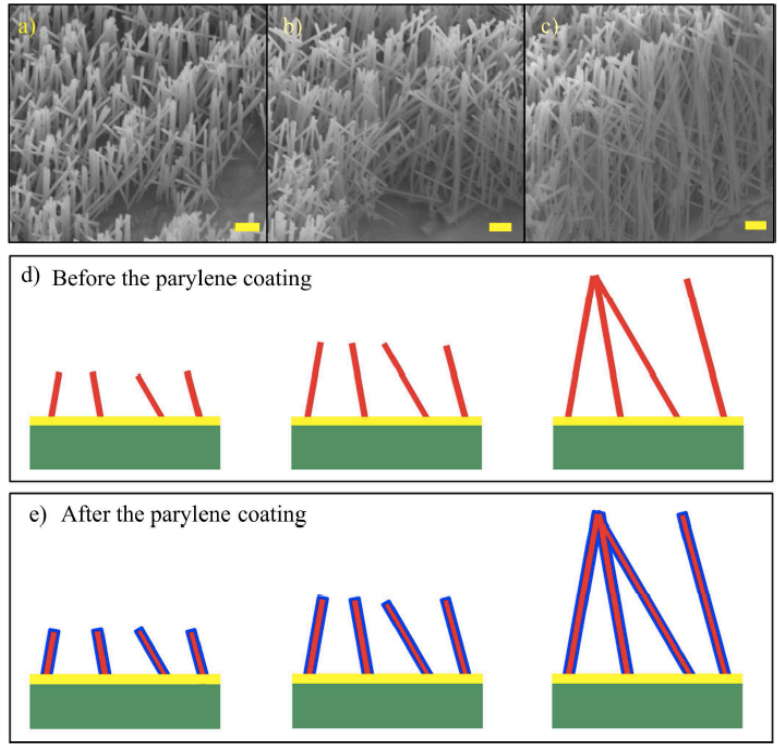

**Figure 12.** (a-c) Side-view SEM images of copper nanowire arrays before parylene coating with the length of 5, 10 and 20 μm. (d) Illustration of nanowire arrays before the parylene coating with the length of 5, 10 and 20 μm. (e) Illustration of nanowire arrays after the parylene coating with the length of 5, 10 and 20 μm. The scale bar is 1 μm. [12]

Besides the nanowire length, the adhesive strength and electrical resistance are also affected by the parylene thickness. The samples with the same nanowire length (10 μm) and the same preload (78.02 N/cm$^2$) are used in the test. As can be seen in Figure 11(c), the adhesive properties strongly depend on the thickness of the parylene shell. Specifically, weak adhesive strengths (~0.99 N/cm$^2$ in shear and ~0.57 N/cm$^2$ in normal directions) are obtained from the pristine copper nanowires. The adhesive strength is dramatically enhanced by the application of the parylene shell. When the thickness of the parylene shell is 150 nm, the maximum adhesive strengths (~24.97 N/cm$^2$ in shear and ~10.82 N/cm$^2$ in normal directions) are obtained. This significant enhancement in adhesion is attributed to the higher surface compliance of the parylene shell, enabling conformal contact with increased contact area between the interpenetrating nanowires. [8] When the thickness of parylene shell further increases, the adhesive strengths decrease. This trend is attributed to the higher filling factor for thicker parylene shells [Figure 5(b)-(d)]. When the thickness of the parylene shell increases to 250 nm, almost no spare space exists between the neighboring nanowires. [12] Hence, the interconnected mode changes from " wire-wire" to "tip-tip" when the thickness of paryelene shell increases, which results in the reduction of adhesive strengths. The electrical properties of nanowire surface fasteners are also affected by the parylene shell thickness. It can be seen from Figure 11(d) that larger parylene shell thickness results in larger electrical resistance of the nanowire surface fasteners. This trend is attributed to the poor electrical conductivity of parylene. To examine the effect of preload on the adhesive and electrical properties of copper/parylene core/shell nanowire surface fastener, two nanowire surface fastener samples were brought into interconnection at a preload of 39.01, 78.02, and 156.04 N/cm$^2$. A monotonic increase in the normal adhesive strength and decrease in electrical resistance are observed with the increase of preload force [Figure 11(e), (f)]. This phenomenon is just as expected because the higher preload force leads to a larger contact area between the nanowires. However, no increase in the shear adhesion was observed when the preload increases from 78.02 to 156.04 N/cm$^2$. This phenomenon is attributed to the poor adhesion of electrodeposited copper nanowires on the Au seed layer. [12]

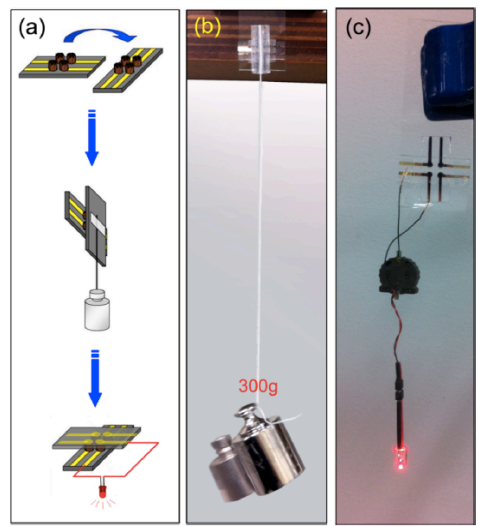

**Figure 13.** (a) Schematic of this room-temperature electrical bonding technique. (b) Photo showing a weight of 300 g hanging on the interconnected copper nanowire surface fasteners. (c) Light-emitting diode suspended by the nanowire surface fasteners to show electrical conductivity. [12]

Figure 13(a) shows the schematic of copper/parylene core/shell nanowire surface fasteners. An example of the strong bonding achieved is shown in Figure 13(b) in which the copper/parylene core/shell nanowire surface fasteners with a surface area of ~3.14×4 mm$^2$ enables 300 g of weight to be hung without failure in shear direction. As shown in Figure 13(c), the red light from the light-emitting diode shows that the core/ shell nanowire surface fasteners are conductive.

**6.2 Performances of copper/polystyrene nanowire surface fasteners**

To demonstrate the importance of the polystyrene shell, we systematically studied the adhesion strength as a function of polystyrene thickness. The samples with the same nanowire length (~40 μm) and the same preload (9.8 N) are used in the test. As can be seen in Figure 14(a), the adhesive properties strongly depend on the thickness of the polystyrene shell. Specifically, weak adhesion strengths (~1.25 N/cm$^2$ in shear and~0.76 N/cm$^2$ in normal directions) are obtained from the pristine copper nanowires. The adhesion strength is dramatically enhanced by the application of polystyrene shell. When the thickness of the polystyrene shell is 18 nm, the maximum adhesion strengths (~44.42 N/cm$^2$ in shear and ~21.43 N/cm$^2$ in normal) are obtained. This drastic enhancement of the adhesion strength with polystyrene shell thickness can be explained by an increase in contact width between the engaged nanowire arrays. [8] A decrease in adhesion strength is observed for the thickness of the polystyrene shell is larger than 18 nm. This trend is attributed to decrease the density of Cu core because of thicker polystyrene shells. When the thickness of the polystyrene shell increases to 24 nm, the polystyrene nanotube becomes more hydrophobic which prevents the electrolyte getting into the polystyrene nanotube and then decreasing the density of Cu core. [70] The electrical properties of copper/polystyrene core/shell nanowire surface fastener are also affected by the polystyrene shell thickness. It can be seen from Figure 14(b) that larger polystyrene shell thickness results in larger electrical resistance of the nanowire surface fastener. This trend is attributed to the poor electrical conductivity of polystyrene. [11]

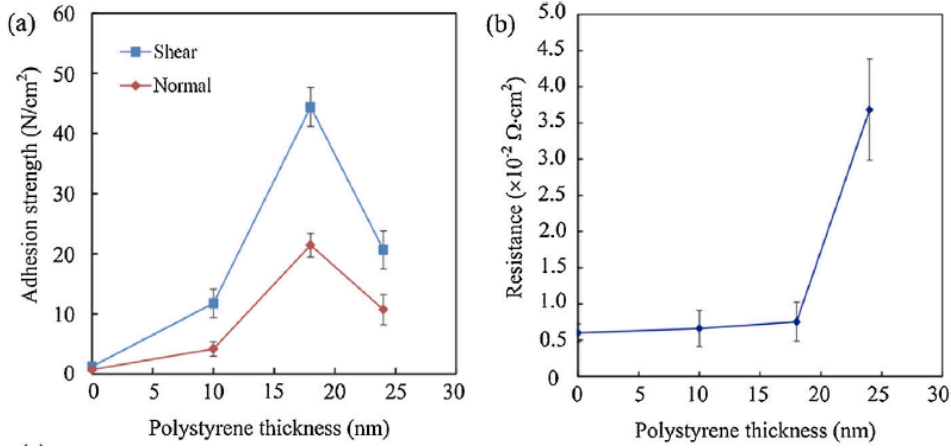

**Figure 14.** (a) Adhesion strength and (b) electrical resistance of nanowire surface fastener as a function of polystyrene thickness. The preload is 9.8 N and the length of nanowire array is ~40 μm. [11]

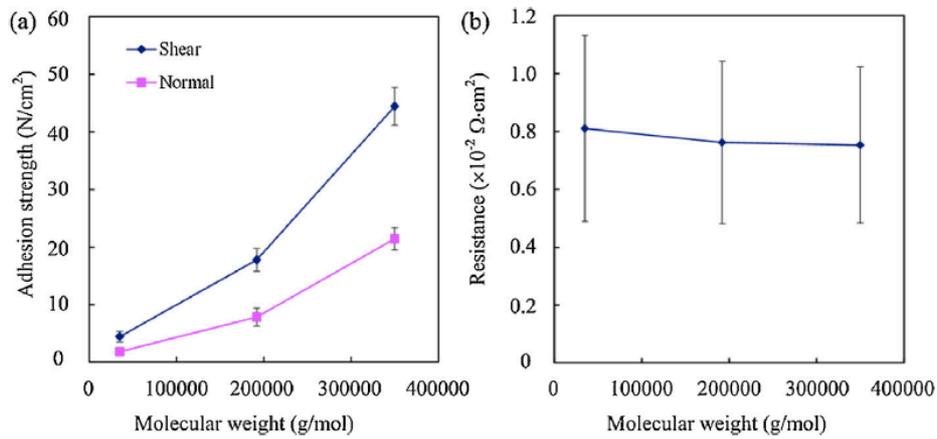

**Figure 15.** (a) Adhesion strength and (b) electrical resistance of nanowire surface fastener as a function of the molecular weight of polystyrene. The preload is 9.8 N and the thickness of polystyrene shell is ~18 nm. [11]

Besides the polystyrene shell thickness, the adhesion strength is also affected by the molecular weight of polystyrene shell. The samples with the same shell thickness (18 nm) and the same preload (9.8 N) are used in the test. As can be seen in Figure 15(a), a monotonic increase in the adhesion strength is observed with the molecular weight of polystyrene. Specifically, the shear adhesion strength increases from 4.64 to 44.42 N/cm$^2$ as the molecular weight of polystyrene is increased from 35,000 to 350,000 g/mol. The adhesion energy of polystyrene increases with an increase in the molecular weight, thus enhancing the adhesion forces between the core/shell nanowires. [71] On the other hand, the changes of the electrical resistance for the nanowire surface fastener were rather small when the molecular weight of polystyrene varied [Figure 15(b)]. The reason is due to that polystyrene shells reach dielectric breakdown and become conductive at this thickness, which is no relationship with the molecular weight of polystyrene shells. [11]

Compare with the copper/parylene core/shell nanowire surface fastener, [12] the copper/polystyrene core/shell nanowire surface fastener tends to achieve higher adhesion strength. Different with parylene, polystyrene is a kind of solvent-welding plastic. Plastic solvent-welding is a technique which uses a solvent to partially liquefy plastic along the joint and allows the joint to solidify causing a permanent chemical weld. Besides the van der Waals force, the chemical force has a big contribution to the adhesion strength. That is the reason that polystyrene shell has a better performance than parylene shell for nanowire surface fastener. Moreover, the polymer shell thickness of copper/polystyrene core/shell nanowire surface fastener is much thinner than that of copper/parylene core/shell nanowire surface fastener, [12] which leads to a good electrical conductivity of copper/polystyrene core/shell nanowire surface fastener. On the other hand, the copper/polystyrene core/shell nanowire surface fastener was fabricated by synthesizing copper nanowires into polystyrene nanotubes, which may easy to form homogeneous polymer shell along the whole length of copper nanowires by comparing with copper/parylene core/shell nanowire surface fastener which was formed by coating parylene shell on core copper nanowires. [11]

### 6.3 Performances of CNT-copper/parylene nanowire surface fasteners

To characterize the performance of a CNT-Cu/parylene nanowire array fastener, we systematically measured the macroscopic shear and normal adhesion strength as a function of the CNT array length [Figure 16(a)]. This revealed that the adhesion strength of a CNT-Cu/parylene nanowire array fastener steadily increases with the length of the CNT array. For example, the shear adhesion strength increases almost linearly from 20.29 to 50.72 N/cm$^2$ when the length of the CNT array increases from 35 to 120 μm. The maximum shear adhesion strength obtained of 50.72 N/cm$^2$ is comparable to that achievable with CNT-based gecko adhesives, [69, 72] and is six-times higher than that of a metallic nanowire fastener. [1, 9, 73] The corresponding normal adhesion strength also increases from 14.82 to 28.48 N/cm$^2$ over the same range of CNT lengths, which is approximately five-times higher than that of a metallic nanowire fastener. [1, 9] Unlike gecko adhesives, the CNT-Cu/parylene nanowire array fastener produces both high shear and normal adhesion strengths simultaneously, which is important for the practical application of electrical fasteners.

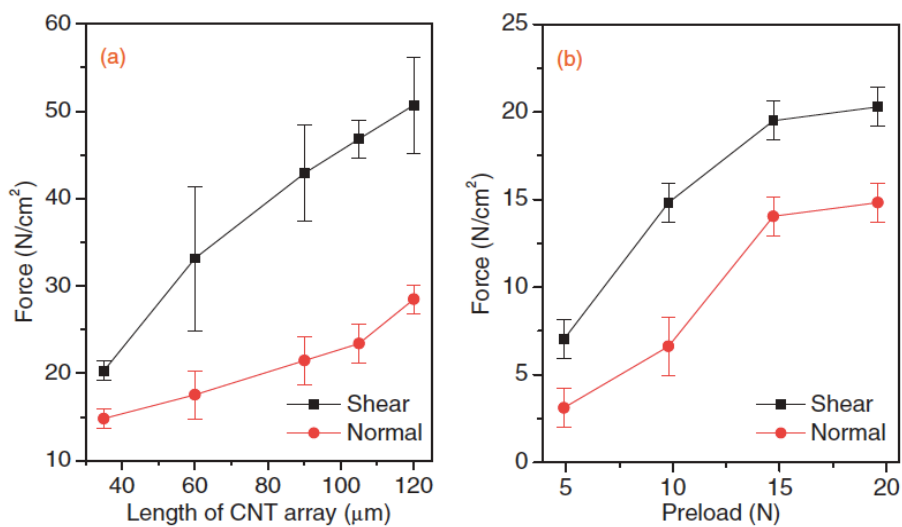

**Figure 16.** (a) Shear and normal adhesion force of a CNT-Cu/parylene nanowire array fastener as a function of the CNT array length. The preload force used to engage the fastener was 19.6 N. (b) Preload force-dependent adhesion force of a CNT-Cu/parylene nanowire array fastener. The errors represent standard errors calculated from 4 measurements. [2]

Besides the height of the CNT array, the adhesion strength of CNT-Cu/parylene nanowire array fasteners is also affected by the preload force applied to engage the fastener. As shown in Figure 16(b), the adhesion strength of a CNT-Cu/parylene nanowire array fastener increases with the applied preload force. Specifically, it increases from 7.02 to 20.29 N/cm$^2$ as the preload force is increased from 4.9 to 19.6 N. The interpenetration depth between CNTs and Cu/parylene nanowires increases with an increase in the applied preload force; thus enhancing the contact area and van der Waals interactions between the CNTs and Cu/parylene nanowires. [74, 75]

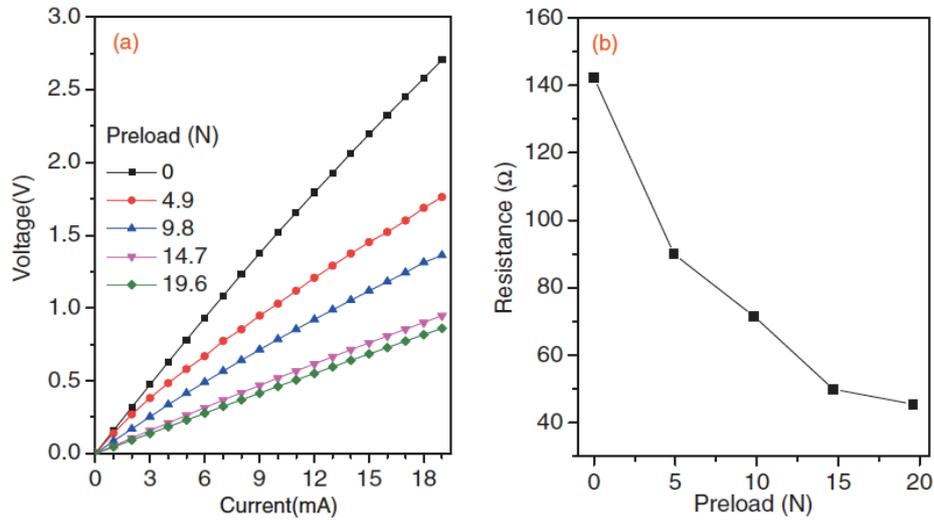

**Figure 17.** (a) *I-V* curves of CNT-Cu/parylene array fastener under different preload force. (b) Measured resistance as a function of preload force. [2]

As shown in Figure 17(a), the electrical resistance of the CNT-Cu/parylene nanowire array fastener was investigated. The solid lines in the figure are obtained through linear fitting of the measured values. The resistance of the fastener was extracted from the *I-V* curves [Figure 17(b)]. [8] As shown in Figure 17(a), the fastener exhibits ohmic behavior over the entire range of measurement; however, the resistance of the fastener decreases with an increase in the preload force. For example, the resistance decreases from 142.3 to 45.4 Ω when the preload force is increased from 0 to 19.6 N, which indicates that the CNT-Cu/parylene nanowire array fastener has good electrical conductivity. It should also be noted that the resistance of the CNT-Cu/parylene nanowire array fastener was almost constant for the first 20 cycles of attachment-detachment tests. Since an increase in preload force increases the interpenetration depth and connection between CNTs and Cu/parylene nanowires, the electrical performance of the fastener is also improved. [2]

# 7. Summary

The state-of-the-art progress on the use of composite nanowires for room-temperature mechanical and electrical bonding has been summarized in this chapter. The fabrication of Cu/parylene and Cu/polystyrene nanowires were described by using AAO and PC membrane as template. Meanwhile, the fabrication of CNTs array was also summarized. For copper/parylene core/shell nanowire array, both strong bonding and small electrical resistance were achieved at room temperature for the copper/parylene core/shell nanowire surface fastener. It is unique that this electrical surface fastener exhibits high macroscopic adhesion strength (~25 N/cm$^2$) and low electrical resistance (~4.22×10$^{-2}$ $\Omega \cdot cm^2$). We also developed an electrical surface fastener with strong adhesion based on copper/polystyrene core/shell nanowire arrays. The adhesion strength of this surface fastener could be mediated by the shell thickness and the molecular weight of polystyrene. Uniquely, this electrical surface fastener exhibits high macroscopic adhesion strength (~44.42 N/cm$^2$) and low electrical resistance (~0.75×10$^{-2}$ $\Omega \cdot cm^2$), which means the copper/polystyrene core/shell nanowire surface fastener shows a higher adhesion strength and a lower electrical resistance than the copper/parylene core/shell nanowire surface fastener. Besides, copper/parylene nanowires and a CNT array were also chosen to construct a room-temperature electrical fastener. The adhesion strength of this fastener was found to increase with an increase in the length of the CNT arrays. The shear adhesion strength (50.72 N/cm$^2$) of the CNT-Cu/parylene nanowire array fastener is six times higher than that of a metallic nanowire fastener. The resistance of the fastener was measured as 45.4 $\Omega$, which indicates that it has good electrical conductivity. Compare with conventional reflow soldering method, the present cold bonding technique can be performed at room temperature, which could improve the process compatibility and component reliability. Furthermore, a type of surface fastener without solder enables easier detachment of the surface mount component from the circuit board, by which recycling of rare metals becomes significantly more convenient.